\documentclass[letter]{aa}  
\usepackage{graphicx}
\usepackage{txfonts}
\begin{document} 

   \title{A cosmic dance at $z\sim3$: Detecting the host galaxies of the dual AGN system LBQS~0302$-$0019 and \textit{Jil} with HAWK-I+GRAAL\thanks{Based on observations collected at the European Organisation for Astronomical Research in the Southern Hemisphere under ESO programme(s) 60.A-9471(A) and 100.A-0134(B).}}
  \titlerunning{Detecting the host galaxies of the dual AGN system LBQS~0302$-$0019 and \textit{Jil}}
   \author{B. Husemann\inst{1}
          \and
          R. Bielby\inst{2}
          \and
          K. Jahnke\inst{1}
          \and
          F. Arrigoni-Battaia\inst{3}
          \and
          G. Worseck\inst{4}          
          \and
          T. Shanks\inst{2}
          \and
          J. Wardlow\inst{2}
          \and
          J. Scholtz\inst{2}
          }

   \institute{Max Planck Institute for Astronomy, K\"onigstuhl 17, D-69117 Heidelberg, Germany, \email{husemann@mpia.de}
         \and 
	    Centre for Extragalactic Astronomy, Durham University, South Road, Durham, DH1 3LE, UK
	 \and
            European Southern Observatory, Karl-Schwarzschild-Str. 2, D-85748 Garching bei M\"unchen, Germany
	 \and
	  Institut f\"ur Physik und Astronomie, Universit\"at Potsdam, Karl-Liebknecht-Str. 24/25, D-14476 Potsdam, Germany  
          }

   \date{}

  \abstract
   {We recently discovered that the luminous radio-quiet QSO LBQS~0302$-$0019 at $z=3.286$ is likely accompanied by an obscured AGN at $20$\,kpc projected distance, which we dubbed \textit{\textit{Jil}}. It represents the tightest candidate obscured/unobscured dual AGN system at $z>3$. To verify the dual AGN scenario we obtained deep $K_s$ band (rest-frame $V$ band) imaging with the VLT/HAWK-I+GRAAL instrument at 0\farcs4 resolution during science verification in January 2018. Indeed, we detect the individual host galaxies of the QSO and \textit{Jil} with estimated stellar masses of $\log(M_\star/M_{\odot})=11.4\pm0.5$ and $\log(M_\star/M_{\odot})=10.9\pm0.5$, respectively. Near-IR spectra obtained with VLT-KMOS reveal a clear [\ion{O}{iii}] $\lambda$5007 line detection at the location of \textit{Jil} which does not contribute significantly to the $K_s$ band flux. Both observations therefore corroborate the dual AGN scenario. A comparison to Illustris cosmological simulations suggests a parent halo mass of $\log(M_\mathrm{halo}/M_{\odot})=13.2\pm0.5$ for this interacting galaxy system, corresponding to a very massive dark matter halo at that epoch. }

   \maketitle
%

\section{Introduction}
Major mergers were initially thought to be one of the main triggering mechanisms for luminous quasi-stellar objects \citep[QSO, e.g.][]{Sanders:1988b,Canalizo:2001,Hopkins:2005}. While major mergers certainly promote gas fueling towards the centers of galaxies in merging systems \citep[e.g.][]{Matteo:2005,Springel:2005}, it is currently heavily debated whether this is really the dominant mode for fueling supermassive black holes (SMBHs) and triggering the most luminous QSOs \citep[e.g.][]{Cisternas:2011,Kocevski:2012,Mechtley:2016}. For QSOs at $z>6$, a [\ion{C}{ii}] survey with ALMA revealed that luminous QSOs exhibit an excess in the number counts of massive companion galaxies within $<$100kpc \citep{Decarli:2017}, suggesting that dense environments and interactions might play an important role in the rapid evolution of the first SMBH systems in the Universe. However, for these systems, only one of the expected SMBHs in the merging galaxies is usually seen to be active. 

Recently, we discovered with MUSE at the VLT that the luminous radio-quiet QSO LBQS~0302$-$0019 is accompanied by a luminous \ion{He}{ii} emitter, dubbed \textit{Jil}, about $2\farcs9$ ($20$\,kpc) away \citep{Husemann:2018}. The emission of \textit{Jil} is best-explained by an embedded obscured active galactic nucleus (AGN), so that the system represents the tightest unobscured/obscured AGN pair at $z>3$. \citet{Frey:2018} analyzed archival Very Large Array radio images which reveal radio emission at the location of LBQS~0302$-$0019 but not at the position of \textit{Jil}. This is expected given the expected AGN luminosity ratio and depth of the radio data. The current radio data therefore do not provide additional constraints on the nature of \textit{Jil} and high-resolution X-ray observations with \textit{Chandra} have not been obtained so far.

In this Letter we present $K$-band spectroscopy and adaptive-optics assisted imaging confirming the presence of a massive host galaxy at the location of \textit{Jil} as a necessary requirement for the dual AGN scenario. Furthermore, we estimate the associated halo mass of this system based on one of the current hydro-dynamical numerical simulations. 

We adopt a flat cosmology with $\Omega_\mathrm{m}=0.3$, $\Omega_\Lambda=0.7$, and $H_0=70$\,km\,s$^{-1}$\,Mpc$^{-1}$. The physical scale at $z=3.286$ is $7.48\,\mathrm{kpc}\,\mathrm{arcsec}^{-1}$ and magnitudes are given in the Vega system.
 
 \begin{figure*}
  \centering
  \includegraphics[width=0.95\textwidth]{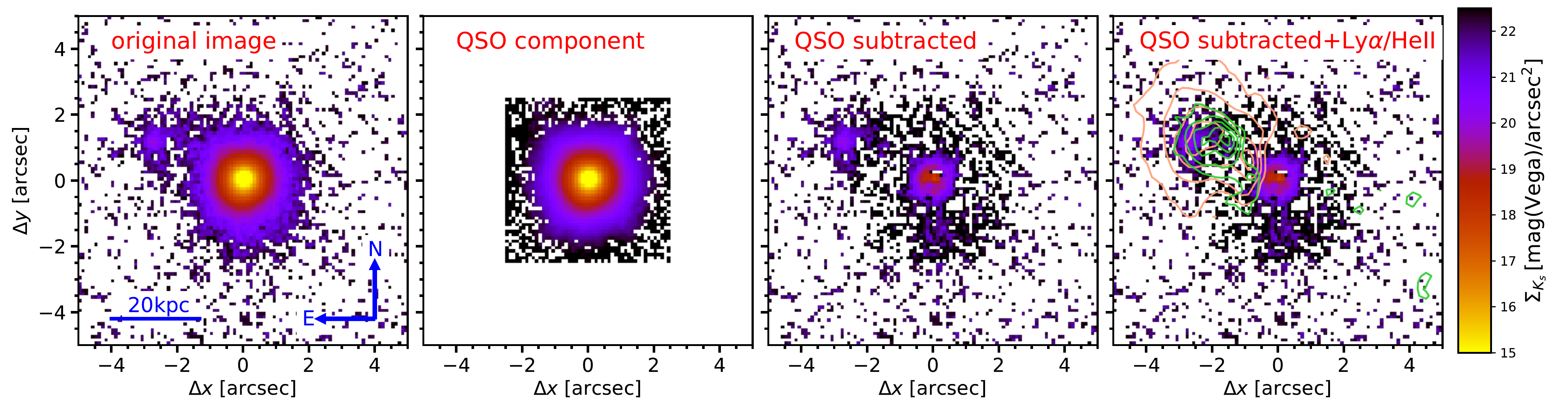}\vspace*{-2mm}
 \caption{$K_s$ band image from VLT/HAWK-I of LBQS~0302$-$0019 and \textit{Jil} with a spatial resolution of 0\farcs4 (full width at half maximum, FWHM). From left to right we present a) the original $K_s$ band image, b) the PSF taken from a nearby star properly scaled to the QSO based on the best-fit GALFIT model, c)  the residual image after QSO component subtraction and d) the residual image with overplotted Ly$\alpha$ contours (red) at $0.2,0.4,0.6,0.8\times10^{-16}\,\mathrm{erg}\,\mathrm{s}^{-1}\mathrm{cm}^{-2}\mathrm{arcsec}^{-2}$ and \ion{He}{ii} $\lambda1640$ contours (green) at $0.4,0.6,0.8,1.0,1.2\times10^{-17}\,\mathrm{erg}\,\mathrm{s}^{-1}\mathrm{cm}^{-2}\mathrm{arcsec}^{-2}$ from the VLT-MUSE observations \citep{Husemann:2018} with a spatial resolution of 1\arcsec\ (FWHM). Strong PSF residuals from the QSO in the Ly$\alpha$ and \ion{He}{ii} contours have been masked out for clarity.}\label{fig:HAWKI}
\end{figure*}
 
\section{Observations and data reduction}
\subsection{VLT/HAWK-I+GRAAL $K_s$ band imaging}

We targeted this dual AGN system during the science verification (SV) with the GRound-layer Adaptive optics Assisted by Laser instrument \citep[GRAAL,][]{Paufique:2010}, which provides a seeing enhancer for the wide-field near-infrared imager HAWK-I \citep{Casali:2006} at the Very Large Telescope (VLT). HAWK-I covers a $7\farcm5\times7\farcm5$ field-of-view (FoV) using an array of $2\times2$ Hawaii-2RG detectors with a $15\arcsec$ gap between the four quadrants. We observed LBQS~0302$-$0019 in the $K_s$ band during the SV run from 2--5 January 2018. The observations were split into 2 separate observing blocks consisting of 12 dithered pointings with $20\times10$\,s exposures each. The QSO was centered in quadrants 1 and 3 of the detector array in the respective observing blocks, which amounts to 4800\,s on source exposure time. 

The data were reduced with the standard ESO pipeline for HAWK-I. The photometric zero-point for the combined observations were determined through aperture photometry of two bright 2MASS stars in the common field of the two pointings. We consider that the photometric zero-point ($m_{K_s,0}=22.73$\,mag) estimated this way has an intrinsic uncertainty of 0.1\,mag. An image cutout of the dual AGN system region is shown in Fig.~\ref{fig:HAWKI} and clearly reveals a prominent continuum source right at the expected location of \textit{Jil} at a spatial resolution of 0\farcs4 (FWHM).

\subsection{KMOS NIR integral-field spectroscopy}
We also observed the dual AGN system with the K-band Multi Object Spectrograph \citep[KMOS,][]{Sharples:2013}. The KMOS data were taken on 27th of January 2018 as part of a back-up program of the VLT LBG Redshift Survey \citep[VLRS;][]{Bielby:2013,Bielby:2017}, intended to fill gaps in the guaranteed observing time due to pointing restrictions, sub-optimal sky conditions and gaps in the distribution of primary target fields on the sky.

KMOS is a NIR multi-object integral field unit (IFU) instrument mounted on VLT UT4. It consists of 24 individual $2\farcs8\times2\farcs8$ IFUs (with pixel scales of 0\farcs2). Only two of the IFUs were used to cover the dual AGN system. For these observations, KMOS was operated with the $HK$ grism in place, providing wavelength coverage from 1.484~$\mathrm{\mu}$m to 2.442~$\mathrm{\mu}$m with spectral resolution ranging from $R\sim1500$ to $R\sim2500$ correspondingly.

Observations were taken in nod-to-sky mode with an ABAABAAB pattern, where A and B represent the two nod positions. Given the proximity of \textit{Jil} to LBQS~0302$-$0019, we were unable to place IFUs on the two targets simultaneously and so \textit{Jil} was targeted during nod position B, whilst LBQS~0302$-$0019 was targeted in position A. In both cases, the IFUs nodded to selected empty sky locations in their respective `off' nod positions, to facilitate sky removal from the science exposures. The two IFUs were positioned to provide some overlap in their coverage, resulting in overlaps of $\approx0\farcs8$ in R.A. and $\approx1\farcs8$ in declination. Each nod was observed for 600s leading to total integration times on LBQS 0302$-$0019 and \textit{Jil} of 3000s and 1800s, respectively.

The data was reduced using {\sc esorex} with the standard ESO pipeline recipes \citep{Davies:2013}, incorporating dark and flat frame subtraction, wavelength calibration, illumination correction, standard star flux calibration, and the overall processing and stacking. The final image quality of the cube is 0\farcs65 (FWHM).  We used the KMOS {\sc sky-tweak} routine to optimally remove the NIR sky lines and  applied the Zurich Atmosphere Purge \citep[{\sc zap};][]{Soto:2016} code on the final cube to further suppress  sky line residuals.

\section{Results}
\subsection{QSO subtraction}
Given the brightness of the QSO LBQS~0302$-$0019, it is crucial to subtract the QSO light to properly resolve the host galaxies of the QSO and \textit{Jil} in the HAWK-I data. To estimate the QSO contribution we firstly created an empirical point-spread function (PSF) from the nearby star 2MASS~J03044733$-$0007499 ($m_{K_s}=13.46\pm0.04$\,mag) which is just 40\arcsec\ away from the QSO.  In a second step, we used GALFIT \citep[v3,][]{Peng:2010} to model the data as a superposition of a single Sersi\'c profile for each of the two galaxies and a point-source for the QSO.  During the fitting, we fixed the Sersi\'c index to $n=1$ to avoid nonphysically large indices, which are caused by the extreme brightness ratio for the QSO host galaxy and the low spatial resolution of 2.8\,kpc (FWHM) compared to the expected galaxy size. The resulting model and QSO subtracted image are shown in Fig.~\ref{fig:HAWKI}, from which we infer $m_{K_s}=19.2\pm0.1$\,mag for the QSO host galaxy and $m_{K_s}=20.6\pm0.2$\,mag for \textit{Jil}.

\begin{figure}
\centering
\includegraphics[width=0.42\textwidth]{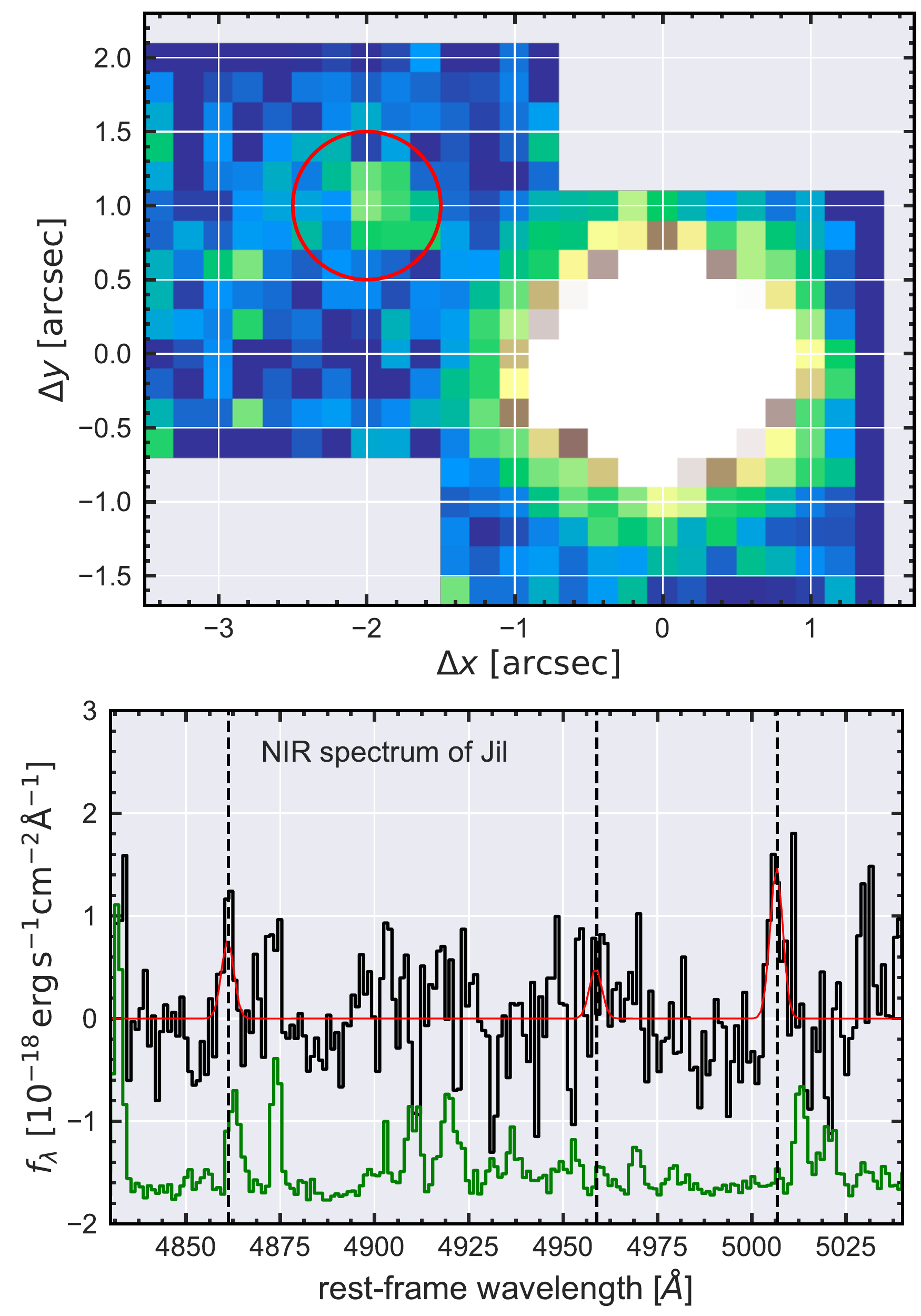}
 \caption{\textit{Upper panel:} Narrow-band image (25\AA\ wide in the observed frame) centered on the redshifted [\ion{O}{iii}] $\lambda5007$ extracted from KMOS datacube. The bright QSO dominates the emission, but a weak source is detected at the location of \textit{Jil}. \textit{Lower panel:} Aperture spectrum (black line) within 0\farcs5 radius centered on \textit{Jil} zoomed into the wavelength range covering the redshifted H$\beta$ and [\ion{O}{iii}] $\lambda\lambda4960,5007$ lines (vertical dashed lines). The green line represent the error spectrum (offset by $-2\times10^{-18}$ for readability) to highlight the position of sky lines. The best-fit model with fixed [\ion{O}{iii}] doublet ratio is shown as the red line.}\label{fig:KMOS}
\end{figure}

\subsection{Host morphology}
The HAWK-I $K_s$ image (Fig.~\ref{fig:HAWKI}c) reveals that the host galaxy of \textit{Jil} appears asymmetric in the rest-frame $V$ band with a faint extension towards the west side. It is unclear at the given spatial resolution and depth whether \textit{Jil} is one distorted galaxy or itself a merger of two galaxies. In comparison, the bright knot in the \ion{H}{i} Ly$\alpha$ nebula recovered with MUSE is centred on \textit{Jil} (Fig.~\ref{fig:HAWKI}d) and traces its morphology well considering the more than 2$\times$ lower spatial resolution. The same applies to the \ion{He}{ii} $\lambda$1640\AA\ emission, but it is more compact with a slight preference to the peak shortly west of \textit{Jil}. The clear matching of the highly ionized gas with the presence of a distinct galaxy at the location of \textit{Jil} is consistent with the picture of an obscured AGN at its centre, but it is not yet a definite proof for the presence of an AGN as such.

\subsection{Rest-frame optical nebular emission}
From the KMOS data we re-constructed an [\ion{O}{iii}] $\lambda$5007 narrow-band image at the redshift of the system (Fig.~\ref{fig:KMOS}). Besides the bright emission from the QSO we detect also faint [\ion{O}{iii}] emission with $5\sigma$ significance at the location of \textit{Jil} as expected for an obscured AGN. The observed line flux is $f_{[\ion{O}{iii}]}=(2.5\pm0.5)\times10^{-17}\,\mathrm{ erg}\,\mathrm{cm}^{-2}\,\mathrm{s}^{-1}$ corresponding to $L_{[\ion{O}{iii}]} = (2.4\pm0.5)\times10^{42}\,\mathrm{ erg}\mathrm{s}^{-1}$. Adopting the conversion factor of $L_\mathrm{bol}/L_\mathrm{[\ion{O}{iii}]}\approx3500$ from \citet{Heckman:2004}, we estimate a bolometric luminosity for \textit{Jil} of $L_\mathrm{bol}\sim4.4\times 10^{45}\,\mathrm{ erg}\mathrm{s}^{-1}$ with a systematic uncertainty of about 0.4\,dex due to the scatter in the relation. 

Compared to the bolometric luminosity of $L_\mathrm{bol}=1\times10^{48}\, \mathrm{ erg}\mathrm{s}^{-1}$ for LBQS~0302$-$0019 \citep{Shen:2016}, the [\ion{O}{iii}]-based luminosity of \textit{Jil} is a factor of 100--620 lower. This is slightly higher than our constraints from the \ion{He}{ii} photoionization models \citep{Husemann:2018}, which required a minimum luminosity of a factor 600--1000 fainter than LBQS~0302$-$0018 assuming a distance of 100\,pc of the ionized gas clouds to the obscured AGN. Hence, the discrepancy of the estimates can be easily explained either by a larger distance of the gas clouds from the AGN, or the effect of dust extinction on the \ion{He}{ii} emission line. 

We cannot detect any other emission lines like [\ion{O}{iii}] $\lambda$4960, H$\beta$ or [\ion{O}{ii}] $\lambda\lambda3726,3729$ at the location of \textit{Jil} in the shallow KMOS data. Since [\ion{O}{iii}] $\lambda$5007 is the brightest line in case of AGN-ionization, we expect non-detections for all other lines given the S/N of our data. To estimate the emission-line contribution to the HAWK-I broad-band observations, we assumed an [\ion{O}{iii}] doublet line ratio of 3 \citep{Storey:2000} and  ([\ion{O}{iii}] $\lambda$5007)/H$\beta\sim10$ to create a mock emission-line spectrum for the $K_s$ band. This leads to an expected pure emission-line brightness of $m_{K_s}=29$\,mag (Vega) confirming that contributions from lines can be safely neglected in the HAWK-I $K_s$ band.
\begin{figure}
\centering
 \includegraphics[width=0.44\textwidth]{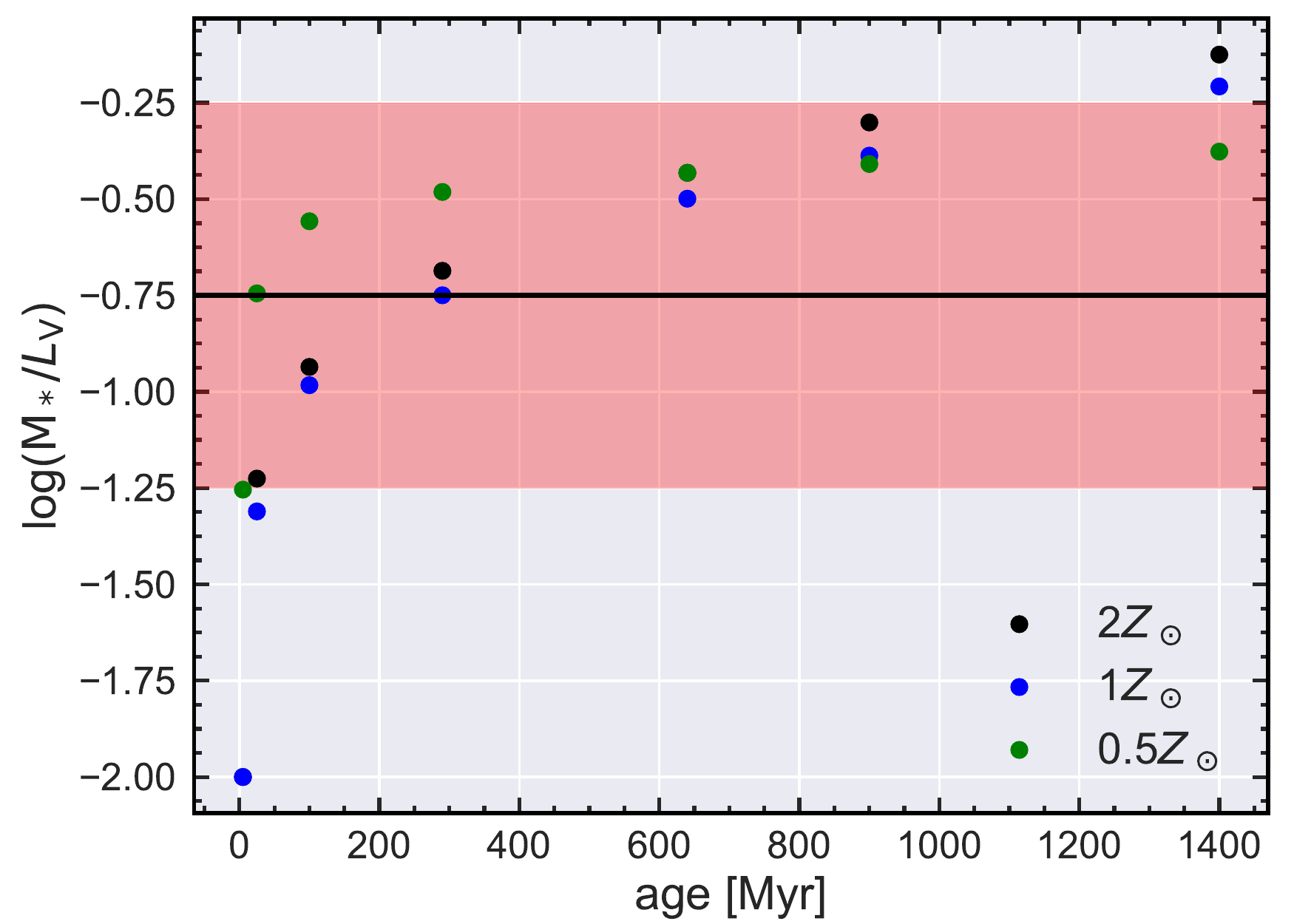}
\caption{Rest-frame $V$ band mass-to-light ratio as a function of stellar population age for three metallicities. For our purpose we adopt a mean mass-to-light ratio of $\log(\mathrm{M_\star}/L_V)=-0.75$  (black line) with an uncertainty of $\pm0.5$ dex (red shaded band).}
\label{fig:mass_ratio}
\end{figure}

\subsection{Stellar masses and halo mass}
At $z=3.3$ the age of the Universe was about $1.8$\,Gyr, which sets a hard boundary for the age of the stellar population. However, the rest-frame $V$ band mass-to-light ratio is still changing by an order of magnitude within the possible range in ages from 100\,Myr to 1.8\,Gyr and metallicity (Fig.~\ref{fig:mass_ratio}), based on the \citet{Bruzual:2003} stellar population models and assuming a Chabrier initial mass function. While the FUV line diagnostics imply sub-solar metallicity \citep{Husemann:2018}, we adopt a mean mass-to-light ratio of $\log(M_\star/L_V)=-0.75\pm0.5$ to be conservative, which leads to stellar masses of $\log(M_\star/M_{\odot})=11.4\pm0.5$ and $\log(M_\star/M_{\odot})=10.9\pm0.5$ for the QSO host and \textit{Jil}, respectively. The uncertainties in the stellar masses are entirely dominated by the uncertainty in the stellar age and the corresponding mass-to-light ratio rather than photometric errors. Given a BH mass of $M_\mathrm{BH}=2.3\times10^9M_\sun$ \citep{Shen:2016}, the inferred host galaxy mass of the QSO is fully consistent with the high-$z$ $M_\mathrm{BH}$--$M_\star$ relation \citep[e.g.][]{Jahnke:2009}.

\begin{figure}
\centering
\includegraphics[width=0.43\textwidth]{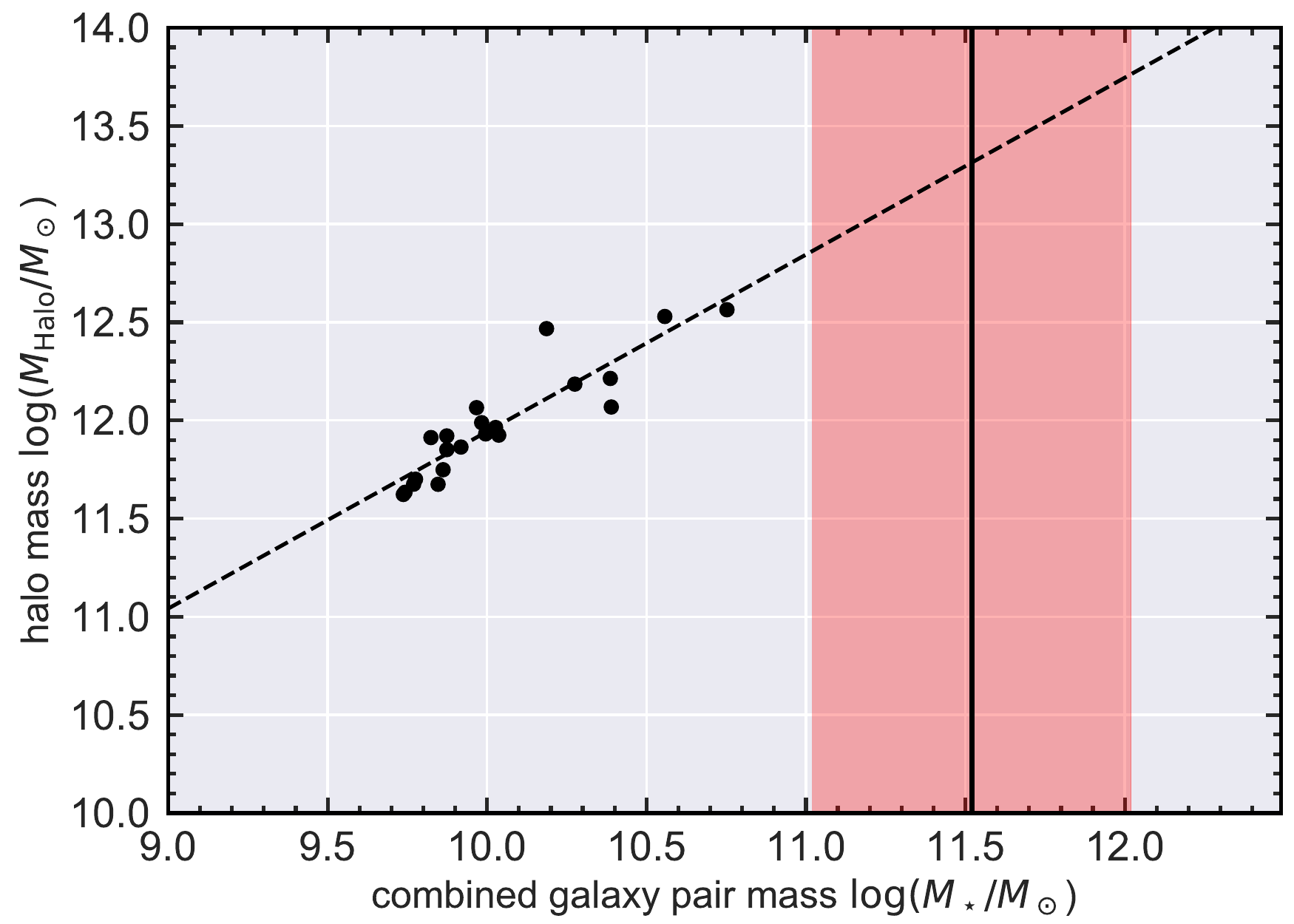}
\caption{Parent halo mass as a function of total stellar mass at $z\sim3.3$ from the Illustris simulation \citep{Vogelsberger:2014} for nearly equal mass $M_1/M_2<10$ galaxy pairs with a separation less than 50\,kpc. The position of our dual AGN pair system is indicated by the vertical black line with uncertainties highlighted by the red shaded area. There is no similar system in Illustris due to the limited volume (106.5\,$\mathrm{Mpc}^3$ comoving), but the extrapolation from lower masses (dashed line) implies a parent halo mass of $12.8<\log(M_\mathrm{halo}/M_\odot)<13.7$.}
\label{fig:illustris}
\end{figure}

To put this system into perspective, we looked at the dark matter halo distribution from the Illustris simulation \citep{Vogelsberger:2014} covering a comoving volume of 106.5\,$\mathrm{Mpc}^3$. We selected halos containing a close pair ($<$50\,kpc) of sub-halos containing stellar masses of $\log(M_\star/M_\sun)>9.0$ and stellar mass ratios $M_1/M_2<10$. At $z\sim3.3$ the Illustris catalog contains $21$ of such systems (Fig.~\ref{fig:illustris}), but none at the total stellar mass we estimated for our dual AGN system. We therefore fit the halo mass as a function of the combined stellar mass with a power-law to extrapolate the observed trend to higher masses.  Given the combined stellar mass for the QSO and \textit{Jil} we find a range in halo masses  of $12.8<\log(M_\mathrm{Halo}/M_\sun)<13.7$, which corresponds to a very massive dark matter halos at that redshift based on predicted halo mass functions \citep[e.g.][]{Watson:2013}.  

\section{Discussion}
Luminous QSOs at high-redshifts are thought to be associated with massive dark matter halos and dense environments  due to the requirement of a rapid BH growth at early cosmic epoch. Observational evidence of high density environments around high-redshift luminous QSOs has been established in several ways. Luminous radio-loud AGN at $1.2<z<3$ have been  found to systematically reside in galaxy overdensities on arcmin scales \citep[e.g.][]{Ivison:2000, Smail:2003, Wylezalek:2013, Rigby:2014,Malavasi:2015,Jones:2015,Silva:2015}. Similar studies for radio-quiet AGN have also revealed galaxy overdensities around them \citep[e.g.][]{Utsumi:2010,Capak:2011,Morselli:2014,Jones:2017}, although contradictory results have been reported \citep{Kikuta:2017, Mazzucchelli:2017}. Furthermore, halo masses of luminous AGN have been estimated through clustering studies, which either found that they reside in overdensities \citep[e.g.][]{Croom:2002,Coil:2009} or the contrary \citep{Coil:2007}. The inconsistencies may be related to the intrinsic properties of QSOs as BH mass has been suggested to correlate most strongly with the halo mass \citep[e.g.][]{Krumpe:2015} or that different galaxy population, such as Ly$\alpha$ emitters, Lyman-break or dusty galaxies, are considered for the clustering analysis. Nevertheless, recent clustering measurements of AGN in the COSMOS field yield a typical halo mass around $10^{13}M_\sun$ at $z\sim3$ \citep{Allevato:2016} in agreement with our results.

Other works focused on the local environment around luminous QSOs. Here, it is striking to see that  luminous QSOs reveal and excess in the number counts of massive star-forming galaxies in their vicinity at $z>6$ \citep{Decarli:2017} and $z\sim4.8$ \citep{Trakthenbrot:2017}. This is in agreement with the notion of a strongly clustered galaxy environment around QSOs at high redshifts \citep[e.g.][]{Garcia-Vergara:2017} and an excess of unobscured dual AGN with $<$40\,kpc separations \citep{Hennawi:2006}. Although the AGN environment is certainly significantly evolving with redshift, luminous AGN, such as LBQS~0302$-$0019, may be signposts  of the most vigorous evolution of galaxies in overdensities at early epochs. 

\section{Conclusion}
Based on deep $K$ band spectroscopy with KMOS and high-resolution imaging with HAWK-I+GRAAL, we have identified the massive ($\log(M_\star/M_{\odot})=10.9\pm0.5$) host galaxy of \textit{Jil}, the obscured companion AGN to LBQS0302-0019 at a projected separation of about 20\,kpc. This clearly supports the obscured AGN nature of \textit{Jil} since the presence of a massive host galaxy implies the existence of a super-massive black hole, potentially powering an AGN. Hence, we expect a direct detection of AGN engine signatures from \textit{Jil} in the radio (core emission), mid-IR (torus) or X-rays (disc corona) with sufficiently deep observations.

The combined stellar mass of both galaxies suggests a very massive parent halo of this intriguing dual AGN system. This is in agreement with observations of the environment around luminous AGN comparable to LBQS~0302$-$0019 at similar or even higher redshifts. It suggests that these luminous AGN are part of and shaped by a vigorous evolutionary phase which might be important to set the properties of massive present-day galaxies.

\begin{acknowledgements}
We thank the anonymous referee for a fast and constructive review. We particularly thank A. van der Wel for helpful discussion about high-redshift galaxy properties and A. Pillepich for an introduction to the Illustris simulations data. RMB, TS and JWL acknowledge the Science and Technology Facilities Council (STFC; grant ST/P000541/1) for support. JLW acknowledges support from an STFC Ernest Rutherford Fellowship (ST/P004784/1). We also thank A. Tiley for facilitating our KMOS observations.
\end{acknowledgements}

\bibliography{references}

\end{document}